\documentclass[epj]{svjour}
\usepackage[sort&compress,square,comma]{natbib}
\usepackage{latexsym}
\usepackage{graphicx}
\usepackage{amsmath}
\usepackage{subfigure}
\begin{document}
\title{Slave-boson theory of the Mott transition in the two-band Hubbard model}
\subtitle{}
\author{A.\ R\"uegg\inst{1}\and M.\ Indergand\inst{1}\and S.\ Pilgram\inst{1}\and  M.\ Sigrist\inst{1}
}                     

\institute{ETH H\"onggerberg, CH-8093 Z\"urich, Switzerland }
\date{Received: date / Revised version: date}
%
\abstract{ We apply the slave-boson approach of Kotliar and Ruckenstein to the
  two-band Hubbard model with an Ising like Hund's rule coupling and bands of
  different widths. On the mean-field level of this approach we investigate
  the Mott transition and observe both separate and joint transitions of the
  two bands depending on the choice of the inter- and intraorbital Coulomb
  interaction parameters. The mean-field calculations allow for a simple
  physical interpretation and can confirm several aspects of previous work.
  Beside the case of two individually half-filled bands we also examine what
  happens if the original metallic bands possess fractional filling either due
  to finite doping or due to a crystal field which relatively shifts the
  atomic energy levels of the two orbitals.  For appropriate values of the
  interaction and of the crystal field we can observe a a band insulating
  state and a ferromagnetic metal.
} 
\maketitle

\section{Introduction}
\label{sec:intro}
The Mott transition in multiorbital systems with several bands gives rise to
complex and intriguing physics.  Multiband systems occur naturally in rare
earth intermetallic compounds and in systems involving transition metals.  In
the former extended conduction electrons and almost localized $f$-electrons
couple through local hybridization and give rise to Kondo and heavy Fermion
physics.  In transition metal oxides, chalcogenides etc. several partially
filled $d$-orbitals are the origin of rather similar electron bands of
different but comparable width. Here the question arises how the interaction
among these orbitals, Coulomb repulsion and Hund's rule coupling, influences
the transition to partially or fully localized degrees of freedom.  What is
the nature of the Mott transition that occurs as the magnitude of the
interactions is increased gradually? In this paper we will be concerned with
these questions.

The two-band Hubbard model with bands of different widths is the simplest
model that captures all the relevant aspects of the Mott transition in
multiorbital systems. In recent years this model was investigated by several
authors \cite{Liebsch:03,Liebsch:04,Koga:04,Koga:04b,Koga:04c,deMedici:05,Ferrero:05,Arita:05,Knecht:05,Liebsch:05,Inaba:05} mainly
in the framework of dynamical mean-field theory (DMFT) and using different
methods to solve the local impurity problem. These calculations have led to
the following understanding of the Mott transition at half filling. Depending
on the exact choice of the intra- and interorbital interaction parameters, one
can observe a sequence of individual Mott transitions in each band or a joint
transition involving both bands simultaneously. The existence of a separate
transition, usually referred to as ``orbital-selective Mott transition''
(OSMT), implies an intermediate phase between the metal and the Mott insulator
where only the narrow band is insulating whereas the wide band still has
metallic properties.  Furthermore, the stability of this intermediate phase
strongly depends on how the Hund's rule coupling is taken into account.

Early studies of the Mott transition in multiorbital systems made by Anisimov
\emph{et al}.\ \cite{Anisimov:02}, Liebsch
\cite{Liebsch:03a, Liebsch:03, Liebsch:04, Liebsch:05} and by Koga and coworkers
\cite{Koga:04, Koga:04b, Koga:04c} and more recent DMFT calculations of de'
Medici \emph{et al}.\ \cite{deMedici:05}, Ferrero \emph{et al}.\ 
\cite{Ferrero:05}, Arita \emph{et al}.\ \cite{Arita:05}, and Knecht \emph{et
  al}.\ \cite{Knecht:05} showed that different impurity solvers capture
different aspects of the Mott transition and can partially lead to different
conclusions. It is therefore desirable to investigate the properties of this
Mott transition within a more analytical theory. We apply the slave-boson
approach of Kotliar and Ruckenstein \cite{KotliarRuckenstein:86} on the
mean-field level, discuss and confirm several aspects of previous work. Our
calculations give reasonable results in a wide range of parameters and allow
in a natural way for a simple physical interpretation. Beside the case of two
individually half-filled bands we examine what happens if the original
metallic bands possess fractional filling either due to finite doping or due
to a crystal field which shifts the atomic energy levels of the two orbitals
relative to each other. In both cases we can observe an OSMT. Due to the
crystal field splitting also a ferromagnetic and a band insulating phase
appear in the phase diagram.

The paper is organized as follows. In Sec.~\ref{sec:model} we introduce the
model. In the rather technical Sec.~\ref{sec:sb} we apply the slave-boson
formalism and its mean-field approximation. The results concerning the Mott
transition are presented and discussed in Sec.~\ref{sec:MITsb} for various
choices of the interaction parameters at half filling, in the presence of a
crystal field and for finite doping. Conclusions and a comparison with
previous results are found in Sec.~\ref{sec:con}.

\section{Model}
\label{sec:model}
We consider the following two-band Hubbard Hamiltonian
\begin{equation}\label{eq:tbhm}
H=\sum_{\alpha\sigma}\sum_{\langle ij\rangle}t_{ij}^{(\alpha)}c_{i\alpha\sigma}^{\dag}c_{j\alpha\sigma}+\hat{V}
\end{equation}
with
\begin{equation}
\label{eq:int}
\hat{V}=U\sum_{i\alpha}\hat{n}_{i\alpha\uparrow}\hat{n}_{i\alpha\downarrow}+\sum_{i\sigma\sigma '}(U'-J\delta_{\sigma\sigma'})\hat{n}_{i1\sigma}\hat{n}_{i2\sigma '}.
\end{equation}
As usual $c_{i\alpha\sigma}^\dag$ ($c_{i\alpha\sigma}$) creates (annihilates)
an electron with spin $\sigma=\uparrow,\downarrow$ and band index $\alpha=1,2$
at the site $i$ and
$\hat{n}_{i\alpha\sigma}=c_{i\alpha\sigma}^{\dag}c_{i\alpha\sigma}$ is the
corresponding occupation number operator. The hopping integral for the orbital
$\alpha$ is denoted by $t_{ij}^{(\alpha)}$. We assume vanishing interorbital
hybridization and that the hopping integrals have different values for the
different orbitals, i.e.\ that the tight-binding bands have different
bandwidths.  The intraband (interband) Coulomb repulsion is denoted by $U$
($U'$) and the Hund's rule coupling by $J$.  In two-band Hubbard models
additional spin-flip and pair-hopping terms are usually included in the Hund's
rule coupling. As shortly discussed in the next section, these terms pose
problems in the slave-boson formalism and we therefore concentrate on the
Ising like Hund's rule coupling in (\ref{eq:int}). Note however that these
terms are not included in Quantum Monte Carlo (QMC) calculations either
\cite{Liebsch:03, Liebsch:04, Knecht:05}.  For a spherically symmetric
screened Coulomb interaction the positive interaction parameters are related
by $U'=U-2J$ \cite{Castellani:78}. The relevant parameter regime is therefore
$U\geq U'$ where the intraorbital repulsion is bigger than the interorbital.

\section{Slave-boson formulation of the two-band Hubbard model}
\label{sec:sb}
\subsection{Slave-boson model}
\label{subsec:sbm}
The treatment of on-site interactions with slave bosons is a well established
method in different fields of strongly correlated electron systems. Kotliar
and Ruckenstein \cite{KotliarRuckenstein:86} introduced this approach for the
(one-band) Hubbard model. By a new functional-integral representation of the
partition function they were able to effectively map the fermionic action on a
bosonic action with local constraints. The simplest saddle-point approximation
of their approach reproduces the results of the Gutzwiller approximation
\cite{BrinkmanRice:70,Gutzwiller:62,Gutzwiller:64}. The slave-boson approach
leads to a novel mean-field theory which is especially useful for examining
the Mott transition.  As the Gutzwiller approximation, the slave-boson
mean-field theory is closely related to Landau's Fermi liquid theory
\cite{Landau:56} since the slave bosons keep track of the other electrons by
measuring the electron occupancy at each atom which leads to a renormalization
of the hopping amplitude and thus to a change of the effective mass.
\begin{figure*}
\centering
\includegraphics[width=0.8\linewidth]{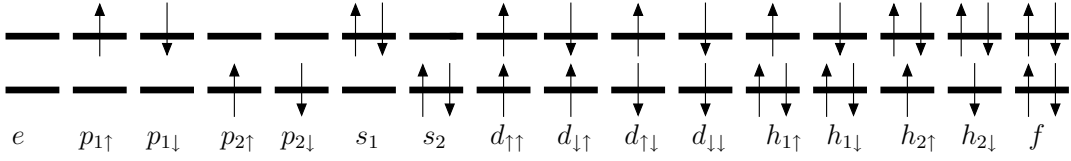}
\caption{The sixteen atomic configurations of the two-band Hubbard model and the
  corresponding slave bosons.} 
\label{fig:config}
\end{figure*}
\begin{table}
\caption{The atomic states in the original model, their corresponding
  slave-boson states as well as the labeling of the mean fields. The site
  index is suppressed, $\alpha=1,2$,  and
  $\bar{\sigma}=\downarrow(\uparrow)$ if $\sigma=\uparrow(\downarrow)$.}
\label{tab:states}
\begin{tabular}{l|rrr}
\hline\noalign{\smallskip}
 &Original model & Slave-boson model &Mean fields \\
\noalign{\smallskip}\hline\noalign{\smallskip}
$|e\rangle$ & $|0\rangle$ &$ e^{\dag}|\mathrm{vac}\rangle$ & $e\equiv \langle e^{(\dag)}\rangle$\\
$|p_{\alpha\sigma}\rangle$ & $c_{\alpha\sigma}^{\dag}|0\rangle$ & $p_{\alpha\sigma}^{\dag}\hat{f}^{\dag}_{\alpha\sigma}|\mathrm{vac}\rangle$ &
$p_{\alpha\sigma}\equiv \langle p_{\alpha\sigma}^{(\dag)}\rangle$\\
$|s_{\alpha}\rangle$ & $c_{\alpha\uparrow}^{\dag}c_{\alpha\downarrow}^{\dag}|0\rangle$ &
$s_{\alpha}^{\dag}\hat{f}_{\alpha\uparrow}^{\dag}\hat{f}_{\alpha\downarrow}^{\dag}|\mathrm{vac}\rangle$
& $s_{\alpha}\equiv \langle s_{\alpha}^{(\dag)}\rangle$\\
$|d_{\sigma\sigma}\rangle$ & $c_{1\sigma}^{\dag}c_{2\sigma}^{\dag}|0\rangle$ &
$d_{\sigma\sigma}^{\dag}\hat{f}_{1\sigma}^{\dag}\hat{f}_{2\sigma}^{\dag}|\mathrm{vac}\rangle$
& $d_{\sigma\sigma}\equiv \langle d_{\sigma\sigma}^{(\dag)}\rangle$\\
$|d_{\sigma\bar{\sigma}}\rangle$ & $c_{1\sigma}^{\dag}c_{2\bar{\sigma}}^{\dag}|0\rangle$ &
$d_{\sigma\bar{\sigma}}^{\dag}\hat{f}_{1\sigma}^{\dag}\hat{f}_{2\bar{\sigma}}^{\dag}|\mathrm{vac}\rangle$
& $d_{\sigma\bar{\sigma}}\equiv \langle
d_{\sigma\bar{\sigma}}^{(\dag)}\rangle$\\
$|h_{1\sigma}\rangle$ & $c_{1\sigma}^{\dag}c_{2\uparrow}^{\dag}c_{2\downarrow}^{\dag}|0\rangle$
&
$h_{1\sigma}^{\dag}\hat{f}_{1\sigma}^{\dag}\hat{f}_{2\uparrow}^{\dag}\hat{f}_{2\downarrow}^{\dag}|\mathrm{vac}\rangle$
& $h_{1\sigma}\equiv \langle h_{1\sigma}^{(\dag)}\rangle$\\
$|h_{2\sigma}\rangle$ & $c_{1\uparrow}^{\dag}c_{1\downarrow}^{\dag}c_{2\sigma}^{\dag}|0\rangle$
&
$h_{2\sigma}^{\dag}\hat{f}_{1\uparrow}^{\dag}\hat{f}_{1\downarrow}^{\dag}\hat{f}_{2\sigma}^{\dag}|\mathrm{vac}\rangle$
& $h_{2\sigma}\equiv \langle h_{2\sigma}^{(\dag)}\rangle$\\
$|f\rangle$ & $c_{1\uparrow}^{\dag}c_{1\downarrow}^{\dag}c_{2\uparrow}^{\dag}c_{2\downarrow}^{\dag}|0\rangle$
&
$f^{\dag}\hat{f}_{1\uparrow}^{\dag}\hat{f}_{1\downarrow}^{\dag}\hat{f}_{2\uparrow}^{\dag}\hat{f}_{2\downarrow}^{\dag}|\mathrm{vac}\rangle$
& $f\equiv \langle f^{(\dag)} \rangle$\\
\noalign{\smallskip}\hline
\end{tabular}
\end{table}
Let us first look at one particular lattice site.  The atomic Hilbert space is
16-dimensional and spanned by the local occupation number basis
listed in the first column of Tab.~\ref{tab:states} and sketched in
Fig.~\ref{fig:config}. The essence of the slave-boson approach of Kotliar and
Ruckenstein is to map the original fermionic model to a mixed
fermionic-bosonic model with local constraints by introducing for each atomic
configuration an auxiliary boson
\begin{equation}\label{eq:bosops}
\{e^{(\dag)},p_{\alpha\sigma}^{(\dag)},s_{\alpha}^{(\dag)},d_{\sigma\sigma'}^{(\dag)},h_{\alpha\sigma}^{(\dag)},f^{(\dag)}\}
\end{equation}
where $\alpha=1,2$, $\sigma=\uparrow,\downarrow$. The labeling of the boson
operators is sketched in Fig.~\ref{fig:config}. In the following we denote the
fermionic annihilation (creation) operators in the slave-boson model by
$\hat{f}_{i\alpha\sigma}^{(\dag)}$ to distinguish them from
$c_{i\alpha\sigma}^{(\dag)}$ defined in the purely fermionic model. In the
extended model, the creation of a general slave-boson state is realized by
acting with the bosonic creation operators (\ref{eq:bosops}) and the new
fermionic operators on the vacuum $|\mathrm{vac}\rangle$.  The states which
correspond to the physical atomic states of the original model are listed in
the second column of Tab.~\ref{tab:states}.

The introduction of the bosonic degrees of freedom leads to unphysical states
which are eliminated by local constraints. Summing up all boson occupancy
operators, we define
\begin{eqnarray}
\hat{I}_i&:=&e_i^{\dag}e_i+\sum_{\alpha\sigma}(p_{i\alpha\sigma}^{\dag}p_{i\alpha\sigma}+h_{i\alpha\sigma}^{\dag}h_{i\alpha\sigma})\nonumber\\&
&+\sum_{\alpha}s_{i\alpha}^{\dag}s_{i\alpha}+\sum_{\sigma\sigma'}d_{i\sigma\sigma'}^{\dag}d_{i\sigma\sigma'}+f_i^{\dag}f_i.
\end{eqnarray}
Furthermore we define the operators
\begin{eqnarray}
\hat{Q}_{i1\sigma}&:=&p_{i1\sigma}^{\dag}p_{i1\sigma}+s_{i1}^{\dag}s_{i1}+\sum_{\sigma'}d_{i\sigma\sigma'}^{\dag}d_{i\sigma\sigma'}\nonumber\\
& &
+h_{i1\sigma}^{\dag}h_{i1\sigma}+\sum_{\sigma'}h_{i2\sigma'}^{\dag}h_{i2\sigma'}+f_i^{\dag}f_i,\\
\hat{Q}_{i2\sigma}&:=&p_{i2\sigma}^{\dag}p_{i2\sigma}+s_{i2}^{\dag}s_{i2}+\sum_{\sigma'}d_{i\sigma'\sigma}^{\dag}d_{i\sigma'\sigma}\nonumber\\
& &
+h_{i2\sigma}^{\dag}h_{i2\sigma}+\sum_{\sigma'}h_{i1\sigma'}^{\dag}h_{i1\sigma'}+f_i^{\dag}f_i.
\end{eqnarray}
The physical subspace is given by the local constraints
\begin{eqnarray}
\hat{I}_i-1& \equiv &0,\label{eq:con1}\\
\hat{f}_{i\alpha\sigma}^{\dag}\hat{f}_{i\alpha\sigma}-\hat{Q}_{i\alpha\sigma}&\equiv&0.\label{eq:con2}
\end{eqnarray} 
These constraints ensure that the slave-boson states listed in the second
column of Tab.~\ref{tab:states} form a complete set in the physical local
Hilbertspace of the slave-boson model. The first relation (\ref{eq:con1})
represents the completeness of the boson operators, i.e., the
sixteen states with one boson form a complete set in the local physical
Hilbertspace of the bosons. The operators $\hat{Q}_{i\alpha\sigma}$ count the
number of bosons that correspond to local configurations having an electron
with spin $\sigma$ in the orbital $\alpha$. Therefore, we have to ensure with
the constraint (\ref{eq:con2}) that in the physical subspace the operators
$\hat{Q}_{i\alpha\sigma}$ are identical to the operators
$\hat{f}_{i\alpha\sigma}^{\dag}\hat{f}_{i\alpha\sigma}$.

Using these constraints, the interaction term becomes quadratic in the boson
operators
\begin{eqnarray}\label{eq:vsb}
\hat{V}^{\mathrm{sb}}&=&\sum_i\Big\{U\sum_{\alpha}s_{i\alpha}^{\dag}s_{i\alpha}+(U+2U'-J)\sum_{\alpha\sigma}h_{i\alpha\sigma}^{\dag}h_{i\alpha\sigma}\nonumber\\
&
&+(U'-J)\sum_{\sigma}d_{i\sigma\sigma}^{\dag}d_{i\sigma\sigma}+U'\sum_{\sigma}d_{i\sigma\bar{\sigma}}^{\dag}d_{i\sigma\bar{\sigma}}\nonumber\\
& &+2(U+2U'-J)f_i^{\dag}f_i\Big\}.
\end{eqnarray}
The attempt to include the spin-flip and pair-hopping term in a similar way
fails due to quartic fermion terms in $V^{\mathrm{sb}}$ or in the constraints.
In this case additional approximations are required \cite{deMedici:05}.
Whereas the interaction term has become much simpler the new formulation of
our model implies that the destruction or creation of a physical fermion has
to be accompanied by slave bosons,
\begin{eqnarray*}
c_{i\alpha\sigma}&\rightarrow &
\tilde{z}_{i\alpha\sigma}\hat{f}_{i\alpha\sigma},\\
c_{i\alpha\sigma}^{\dag}&\rightarrow &
\hat{f}_{i\alpha\sigma}^{\dag}\tilde{z}_{i\alpha\sigma}^{\dag},
\end{eqnarray*}
where
\begin{eqnarray}
\tilde{z}_{i\alpha\sigma}&=&(1-\hat{Q}_{i\alpha\sigma})^{-1/2}z_{i\alpha\sigma}\hat{Q}_{i\alpha\sigma}^{-1/2}\label{ztilde},\\
z_{i1\sigma} &=&
e_i^{\dag}p_{i1\sigma}+p_{i1\bar{\sigma}}^{\dag}s_{i1}+p_{i2\sigma}^{\dag}d_{i\sigma\sigma}+p_{i2\bar{\sigma}}^{\dag}d_{i\sigma\bar{\sigma}}\nonumber\\
&
&+s_{i2}^{\dag}h_{i1\sigma}+d_{i\bar{\sigma}\sigma}^{\dag}h_{i2\sigma}+d_{i\bar{\sigma}\bar{\sigma}}^{\dag}h_{i2\bar{\sigma}}+h_{i1\bar{\sigma}}^{\dag}f_i,\nonumber\\
z_{i2\sigma} &=&
e_i^{\dag}p_{i2\sigma}+p_{i2\bar{\sigma}}^{\dag}s_{i2}+p_{i1\sigma}^{\dag}d_{i\sigma\sigma}+p_{i1\bar{\sigma}}^{\dag}d_{i\bar{\sigma}\sigma}\nonumber\\
& & +s_{i1}^{\dag}h_{i2\sigma}+d_{i\sigma\bar{\sigma}}^{\dag}h_{i1\sigma}+d_{i\bar{\sigma}\bar{\sigma}}^{\dag}h_{i1\bar{\sigma}}+h_{i2\bar{\sigma}}^{\dag}f_i\nonumber.
\label{eq:zfact}
\end{eqnarray} 
The ``$z$-operators'' keep track of the environment (bosons) during hopping
processes \cite{KotliarRuckenstein:86}.  The choice of the ``$z$-operators''
is not unique. In fact, as long as the constraints are fulfilled exactly, the
``$z$-operators'' can be modified by any operator which is the identity
operator when restricted to the physical subspace \cite{KotliarRuckenstein:86,
  Dietrich:94}.  However, the mean-field results depend on the choice of these
operators. The choice of Kotliar and Ruckenstein, that we take here
(\ref{ztilde}), reproduces correctly the noninteracting case in the mean-field
approximation \cite{KotliarRuckenstein:86}. The slave-boson Hamiltonian is
then given by
\begin{equation}\label{eq:sbham}
  H^{\mathrm{sb}}=\sum_{ij\alpha\sigma}t_{ij}^{(\alpha)}\hat{f}_{i\alpha\sigma}^{\dag}\tilde{z}_{i\alpha\sigma}^{\dag}\tilde{z}_{j\alpha\sigma}\hat{f}_{j\alpha\sigma}+\hat{V}^{\mathrm{sb}}
\end{equation}
and is fully equivalent to the original Hamiltonian provided the local
constraints (\ref{eq:con1}, \ref{eq:con2}) are handled exactly.  They can be
imposed by site dependent Lagrange multipliers $\lambda_i^{I}$ and
$\lambda_{i\alpha\sigma}^{Q}$.
\subsection{Mean-field approximation}
\label{subsec:mf}
The simplest saddle-point approximation of the grand canonical partition
function $\mathcal{Z}=\mathrm{Tr}[e^{-\beta(H^{\mathrm{sb}}-\mu
  N^{\mathrm{sb}})}\mathcal{P}]$ is equivalent to a mean-field approximation
where the Bose fields and Lagrange multipliers are treated as static and
homogeneous fields.  Thus, this approximation consists essentially in
replacing the creation and annihilation operators of the slave bosons by site
independent c-numbers which can be chosen to be real. The mean fields are
listed in the fourth column of Tab.~\ref{tab:states}. In this approximation,
the constraints are fulfilled only on average and the square of the mean
fields can be interpreted as the probability of finding the corresponding
local configuration at a particular site. The mean-field Hamiltonian with
included averaged constraints can be diagonalized and yields at $T=0$ the
variational ground-state energy (per site)
\begin{equation}
\label{eq:EGtilde}
\tilde{E}_{\mathrm{G}}=\sum_{\alpha\sigma}q_{\alpha\sigma}\bar{\varepsilon}_{\alpha\sigma}+V_{MF}+\lambda_{\alpha}^{I}(I-1)-\sum_{\alpha\sigma}\lambda_{\alpha\sigma}^{Q}(Q_{\alpha\sigma}-n_{\alpha\sigma}).
\end{equation}
By introducing the effective chemical potential
$\mu_{\alpha\sigma}=\mu-\lambda_{\alpha\sigma}^{Q}$ in band $\alpha$ for
fermions with spin $\sigma$, the average kinetic energy per site and band
reads
\begin{equation}
\bar{\varepsilon}_{\alpha}=\bar{\varepsilon}_{\alpha\uparrow}+\bar{\varepsilon}_{\alpha\downarrow}=\sum_{\sigma}\int_{-\infty}^{\mu_{\alpha\sigma}/q_{\alpha\sigma}}d\varepsilon\varepsilon\rho_{\alpha}(\varepsilon).
\end{equation}
Similarly, the density in band $\alpha$ is
\begin{equation}\label{eq:dens}
n_{\alpha}=n_{\alpha\uparrow}+n_{\alpha\downarrow}=\sum_{\sigma}\int_{-\infty}^{\mu_{\alpha\sigma}/q_{\alpha\sigma}}d\varepsilon\rho_{\alpha}(\varepsilon).
\end{equation}
The chemical potential $\mu_{\alpha\sigma}$ has to be determined from
Eq.~(\ref{eq:dens}) for a given density. If both bands are half-filled
separately one finds for example $\mu_{\alpha\sigma}=0$.  The bare density of
state (DOS) per spin in the band $\alpha$ is denoted by $\rho_{\alpha}$ and
the mean-field Coulomb energy per site is
\begin{eqnarray}
V_{\mathrm{MF}}&=&U\left(\sum_{\alpha}(s_{\alpha}^2+\sum_{\sigma}h_{\alpha\sigma}^2)+2f^2\right)\nonumber\\
& & +U'\left(\sum_{\sigma\sigma'}d_{\sigma\sigma'}^2+2\sum_{\alpha\sigma}h_{\alpha\sigma}^2+4f^2\right)\nonumber\\
& & -J\left(\sum_{\sigma}(d_{\sigma\sigma}^2+h_{1\sigma}^2+h_{2\sigma}^2)+2f^2\right).
\end{eqnarray}
The band-renormalization factor $q_{\alpha\sigma}=\tilde{z}_{\alpha\sigma}^2$
can be related to the effective mass of quasiparticles of Landau's Fermi
liquid theory. For quasiparticles in the band $\alpha$ with spin $\sigma$ we
have $q_{\alpha\sigma}^{-1}=m_{\alpha\sigma}^*/m$ \cite{BrinkmanRice:70,
  Vollhardt:84}. The vanishing of $q_{\alpha\sigma}$ therefore indicates the
transition to a localized state.  The mean fields and the Lagrange multipliers
are determined by the stationary point of $\tilde{E}_{\mathrm{G}}$ which is a
saddle point but not a minimum. With the help of the averaged constraints and
of Eq.~(\ref{eq:dens}) we reduce the number of independent variables. The
stationary point of the variational ground-state energy per site
\begin{equation}\label{eq:EG}
E_{\mathrm{G}}=\sum_{\alpha\sigma}q_{\alpha\sigma}\bar{\varepsilon}_{\alpha\sigma}+V_{\mathrm{MF}}
\end{equation}
then becomes a true minimum and can be found numerically in a rather simple way.

\section{Mott transition in the two-band model within slave-boson theory}
\label{sec:MITsb}

In this section we present the results concerning the Mott transition obtained
by numerically minimizing Eq. (\ref{eq:EG}). Unless otherwise stated we always
assume a paramagnetic ground state, i.e.\ 
$n_{\alpha\uparrow}=n_{\alpha\downarrow}$, and a particle-hole symmetric bare
DOS. Consequently, the following conditions are satisfied: $e=f$, $p_{\alpha}\equiv
p_{\alpha\sigma}=h_{\alpha\sigma'}$, $s\equiv s_1=s_2$, $d_0\equiv
d_{\uparrow\downarrow}=d_{\downarrow\uparrow}$ and $d_1\equiv
d_{\uparrow\uparrow}=d_{\downarrow\downarrow}$. This greatly reduces the
computational effort.

If both bands are separately half-filled, the results of the mean-field
calculations do not depend on the exact choice of the bare DOS, as long as it
is particle-hole symmetric. Away from half filling, the mean-field results
slightly depend on the exact choice. For simplicity, we choose throughout this
section for both bands a rectangular DOS and denote its half-width by
$D_{\alpha}$.  The narrow band is always referred to as band~1 and the wider
as band~2. Unless otherwise stated we choose the ratio of the bandwidths such
that\footnote{Originally, this choice was motivated by the fact that in
  $\mathrm{Ca}_{2-x}\mathrm{Sr}_x\mathrm{RuO}_4$ the $d_{xy}$-bandwidth is
  approximately twice the $d_{xz,yz}$-bandwidth. See \cite{Anisimov:02,
    Liebsch:03a}.} $D_1/D_2=1/2$.  Energy is measured in units of the
bandwidth of band 2, i.e.\ $2D_2=1$. We restrict to the relevant parameter
regime $U\geq U'$, where the intraorbital repulsion is bigger than the
interorbital. In Sec.~\ref{subsec:J0} we discuss the $U$-$U'$ phase diagram
for $J=0$ at half filling.  In Sec.~\ref{subsec:D1D2} we focus on the
dependence of the Mott transition on the ratio $D_1/D_2$ for vanishing Hund's
rule coupling and $U=U'$. In Sec.~\ref{subsec:Jn0} we impose the condition
$U'=U-2J$. In this case we also examine the effect of a crystal field and the
influence of finite doping.

\subsection{$\bf{U\geq U'}$ and $\bf{J=0}$}
\label{subsec:J0}
\begin{figure}
\centering \includegraphics[width=0.8\linewidth]{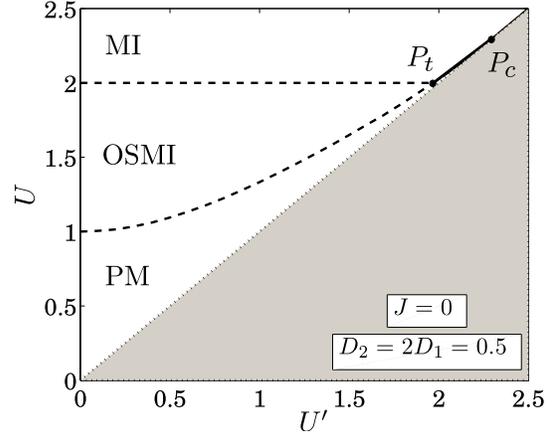}
\caption{Phase diagram at half filling for  $U\geq U'$, $J=0$ and $D_1/D_2=1/2$. Three different phases can be distinguished: a
  paramagnetic metal (PM), a Mott insulator (MI) and in between an orbital
  selective Mott insulator (OSMI). Two second order lines (dashed) merge at
  $P_t$ to a single first order line (solid) which ends in a critical second
  order point $P_c$.}\label{fig:phasesj0}
\end{figure}

\begin{figure}  
  \subfigure[\label{fig:qfacsnbosons0}]{\includegraphics[width=0.49\linewidth]{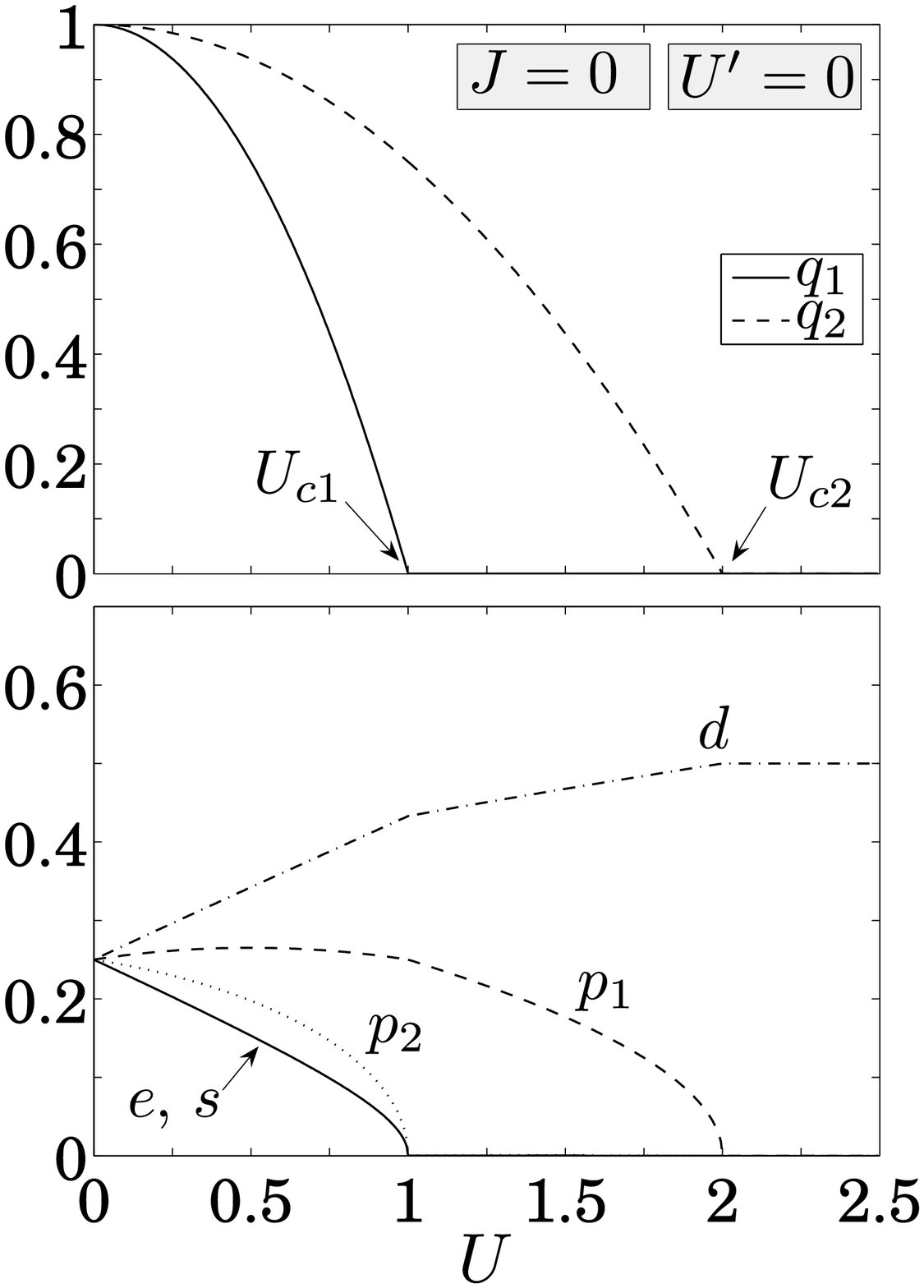}}
  \hfill
  \subfigure[\label{fig:qfacsnbosons1}]{\includegraphics[width=0.49\linewidth]{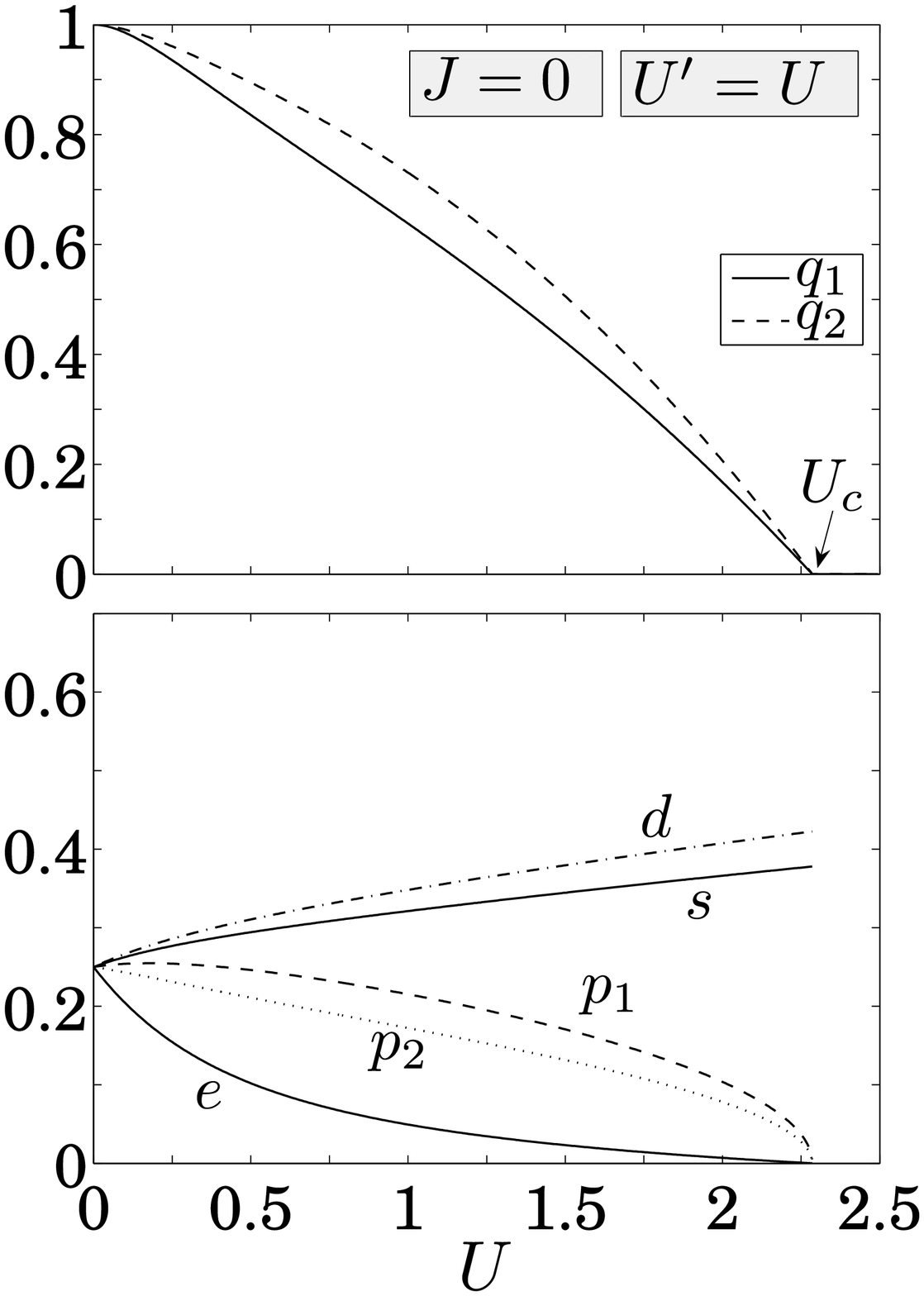}}
\caption{Mean fields of the slave bosons and band-renormalization factors
  $q_{\alpha}$ (a) for $U'=J=0$ and (b) for $U=U'$, $J=0$.}
\end{figure}

The phase diagram at half filling is displayed in Fig.~\ref{fig:phasesj0}. We
can distinguish three different phases: a metallic state (PM), a
Mott-insulating state (MI) and an intermediate state (OSMI) induced by the
OSMT where the localized band 1 coexists with the metallic band 2.
For vanishing Hund's rule coupling, our calculations suggest that the OSMI
phase is bounded by two second order lines (dashed lines in
Fig.~\ref{fig:phasesj0}). They merge to a single first order line at $P_t$
which ends in a critical point $P_c$.  The occurrence of this orbital
selective Mott insulator is not surprising since we have neglected local
spin-spin interactions.  Therefore, if the first band is in a Mott-insulating
state, the electrons in the second band only feel a uniform charge background
(arising from the localized electrons in orbital 1) and the Mott transition in
the broader band occurs at the critical interaction strength
$U_{c2}=8|\bar{\varepsilon}_2^o|=4D_2=2$ which is the value of one independent
band with bandwidth $2D_2=1$ \cite{KotliarRuckenstein:86, BrinkmanRice:70}.
To understand the behavior of the system for general values $0\leq U'\leq U$ it
is instructive to consider first the following two limiting cases:

\emph{i) $U'=0$.} In this case the two bands are independent and the critical
interaction strength of the Mott transition is proportional to the bandwidth
$D_{\alpha}$. In Fig.~\ref{fig:qfacsnbosons0} we have plotted the mean fields
of the slave bosons and the band-renormalization factors. In the
noninteracting system, $U=0$, all configurations are equal likely:
$e=s=p_1=p_2=d=1/4$ where $d\equiv d_0=d_1$.  The vanishing of $q_1$ is
accompanied by the vanishing of $e$, $s$ and $p_2$, whereas $q_2$ becomes
simultaneously zero with $p_1$. In the Mott-insulating phase ($U>U_{c2}$) we
find at each lattice site one of the possible four atomic configurations
represented by the boson $d$ and therefore $d$ reaches $1/2$ at $U=U_{c2}$.

\begin{figure}[h]
  \centering \includegraphics[width=0.5\linewidth]{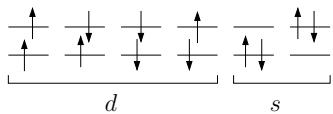}
\caption{For $U=U'$ and $J=0$ there are six degenerate two-electron on-site
  configurations. They are represented by the mean fields $d$ and $s$}
\label{fig:configdeg}
\end{figure}

\emph{ii) $U=U'$.} For this choice the interaction Hamiltonian has an enlarged
symmetry with six degenerate two-electron configurations shown in
Fig.~\ref{fig:configdeg}: four spin configurations with one electron in each
 orbital (represented by $d$) and two configurations with both electrons in one
of the two orbitals (represented by $s$).
This higher symmetry is due to the fact that the Coulomb energy of a local
configuration depends only on the total charge on the atom.  The additional
symmetry in orbital and spin degrees of freedom enlarges the phase space for
charge fluctuations and leads to a stabilization of the metallic phase
\cite{Koga:04}.  In Fig.~\ref{fig:qfacsnbosons1} the mean fields of the slave
bosons as well as the band-renormalization factors are plotted as a function
of $U$. There is a joint Mott transition at the critical interaction strength
$U_c$.  Because of orbital fluctuations and in contrast to the case $U'<U$ not
only configurations represented by $d$ but also configurations represented by
$s$ have a finite probability at $U_c$.  Despite the high symmetry, it
surprisingly turns out that $d\neq s$. The relative strength of the mean
fields in Fig.~\ref{fig:qfacsnbosons1} can be understood as follows. The high
Coulomb energy of a fully occupied local configuration disfavors most strongly
the mean field $e$. The effect of the intraband Coulomb interaction is
stronger in the narrow band and therefore $p_1\geq p_2$ since $p_{1}$ favors
localized behavior in the band~1 and itinerant behavior in the band~2.
Furthermore, the $z$-factors can be approximated by $z_1\approx 2p_1s+4p_2d$
and $z_2\approx 2p_2s+4p_1d$ for high values of $U$. In order to optimize the
hopping in the wide band, $d$ is slightly increased compared to $s$. Note that
above $U_c$ the ratio of $d$ and $s$ is not determined at zero temperature.

The behavior of the system for general values $0<U'<U$ is mostly determined by
the physics of the above discussed two special choices of parameters.

\subsection{$\bf{D_1\ll D_2}$, $\bf{U=U'}$ and $\bf{J=0}$}
\label{subsec:D1D2}
Within our mean-field calculation the existence of a joint transition for
$U=U'$ and $J=0$ depends on the ratio of the bandwidths $D_1/D_2$. It turns
out that for ratios below a critical value $(D_1/D_2)_c$ the mean-field
calculations suggest an OSMT even for $U=U'$. Thus, we recover exactly the
same results as Ferrero \emph{et al}.\ \cite{Ferrero:05} using the Gutzwiller
approximation and de' Medici \emph{et al}.\ \cite{deMedici:05} within their
slave-spin approximation. The critical ratio where an OSMI occurs can be
calculated analytically within the Gutzwiller (or slave-boson) approximation
and is \cite{Ferrero:05} $(D_1/D_2)_c=1/5$. At first sight, the existence of
such a critical ratio seems to be in contradiction with the symmetry argument
given by Koga \emph{et al}.\ \cite{Koga:04}. It states that for vanishing $J$
and $U'=U$ the Mott-Hubbard gap in both bands closes at the same critical
interaction strength, independent of $D_1/D_2$.  However, it does not exclude
the transition into an intermediate phase, where the ``localized'' band is not
fully gapped \cite{deMedici:05}. Indeed, DMFT calculations for $D_1\ll D_2$
\cite{Ferrero:05, deMedici:05} show clearly that the ``localized'' band is not
fully gapped but has spectral weight down to arbitrarily low energies. This
subtle aspect is not captured by the Gutzwiller approximation and related
mean-field theories.

There is however the possibility that the OSMI phase is replaced by an
instability not considered so far. Our mean-field calculation as well as
earlier DMFT calculations did not take into account a possible enlargement of
the unit cell. Below a critical temperature $T_N$ one usually finds
antiferromagnetic long-range order in the Mott insulating phase depending on
the topology of the lattice.  Interestingly for $U'=U$ and $J=0$ spin and
orbital degrees are relevant and it is possible that the OSMI phase is
unstable against an orbitally ordered phase. Whereas for $U'<U$ there is no
tendency toward such an instability it cannot be excluded a priori for $U'=U$.
We discuss now the mechanism which may drive orbital order as an another way
to double the unit cell.

Let us look at the extreme limit $D_1\ll U=U'\ll D_2$ and assume that the lattice is
bipartite. In an adiabatic approximation  the narrow band is localized and
fully dominated by the intraorbital Coulomb repulsion ($D_1/U\approx 0$)
whereas the intraorbital Coulomb interaction in the wide band has negligible
effect ($U/D_2\approx 0$). We look at the following two limiting cases for
the static configuration of the localized (narrow) band and their implications
to the electronic properties of the wide band:
\begin{itemize}
\item[(a)] \emph{Homogeneous charge distribution with exactly one electron per
    orbital.}
  
\item[(b)] \emph{Staggered charge distribution with doubly occupied orbitals
    on one sublattice and empty orbitals on the other sublattice.}
\end{itemize}
In the case (a), the homogeneous charge background contributes an amount $U$
per site to the total energy. In the case (b), the doubling of the unit cell
and the induced rearrangement of electrons in the wide band opens a gap. In
this way the interorbital interaction is reduced ($<U$).
For perfect nesting the wide band is fully gapped and shows insulating
behavior.  On the other hand, the doubly occupied orbitals in the localized
phase cost a fixed amount $U/2$ per site. Thus, there is a competition
between these two effects which can favor an orbitally ordered phase in a
certain parameter range.

In summary, the Mott transition for $U'=U$ is a subtle issue due to the aspect
of possible orbital order and we suggest that this plays a relevant role for
the case $D_1\ll D_2$. Analogous to the spin ordering in the Mott insulator,
the stability of such a phase depends also on details of the band structures.
Taking into account the possibility of a doubling of
the unit cell is an interesting topic for further investigations to be
reported elsewhere.

\subsection{$\bf{U\geq U'}$ and $\bf{U'=U-2J}$}
\label{subsec:Jn0}

We turn back to a given ratio $D_1/D_2=1/2$. From now on we adopt the relation
$U=U'+2UJ$ which is usually used in the discussion of the Mott transition in
the two-band Hubbard model. This relation is valid for a rotationally
symmetric (screened) Coulomb interaction.
\subsubsection{Mott transition at half filling}
\begin{figure}
  \centering \includegraphics[width=0.8\linewidth]{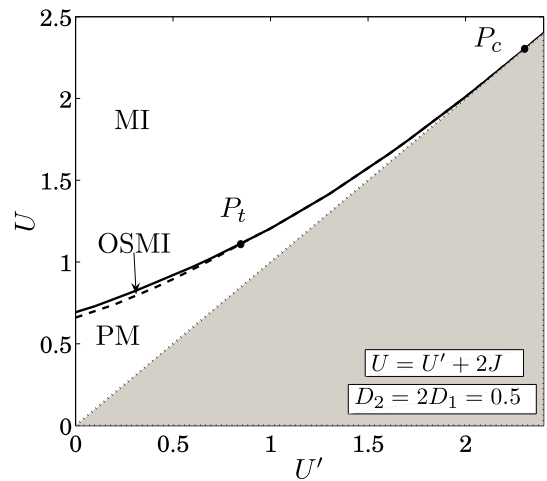}
\caption{Phase diagram at half filling for  $U=U'+2J$ and $D_1/D_2=1/2$. Three different phases can be distinguished: a
  paramagnetic metal (PM), a Mott insulator (MI) and in between an orbital
  selective Mott insulator (OSMI). At $P_t$ the second order line (dashed)
  meets the first order line (solid) which ends in a continuous critical point
  $P_c$.}\label{fig:phases}
\end{figure}

The phase diagram is shown in Fig.~\ref{fig:phases}. Again we can observe the
OSMI phase, but it is limited to a tiny parameter regime. In general the Mott
transition is shifted to smaller values of $U$ since the Ising like Hund's
rule coupling favors localized configurations with parallel spins. For the
same reason the Mott transition in the second band is closer to the one in the
first band because, in contrast to the case $J=0$ discussed in
Sec.~\ref{subsec:J0}, the electrons in the second band not only feel a uniform
charge background but also a localized spin at each lattice site after the gap
for charge excitations in the first band has opened. For small values of $J$
the physics for $U'=U$ becomes important.
The dashed line in Fig.~\ref{fig:phases}, which separates PM-OSMI, is a second
order line whereas the solid line, which separates OSMI-MI and PM-MI, is a
first order line\footnote{Including spin-flip and pair-hopping terms in the
  Hund's rule coupling Koga \emph{et al}.\ \cite{Koga:04} reported two
  successive second-order transitions at $T=0$. For $T>0$ Liebsch
  \cite{Liebsch:05} identified a sequence of two first-order transitions for
  the same model. If spin-flip and pair-hopping terms are omitted he found a
  first-order transition followed by a continuous transition.}. They merge at
$P_t$. The first order line ends in a second order transition point $P_c$
(Fig.~\ref{fig:qfacsnbosons1}).

\begin{figure}
\centering
  \includegraphics[width=1\linewidth]{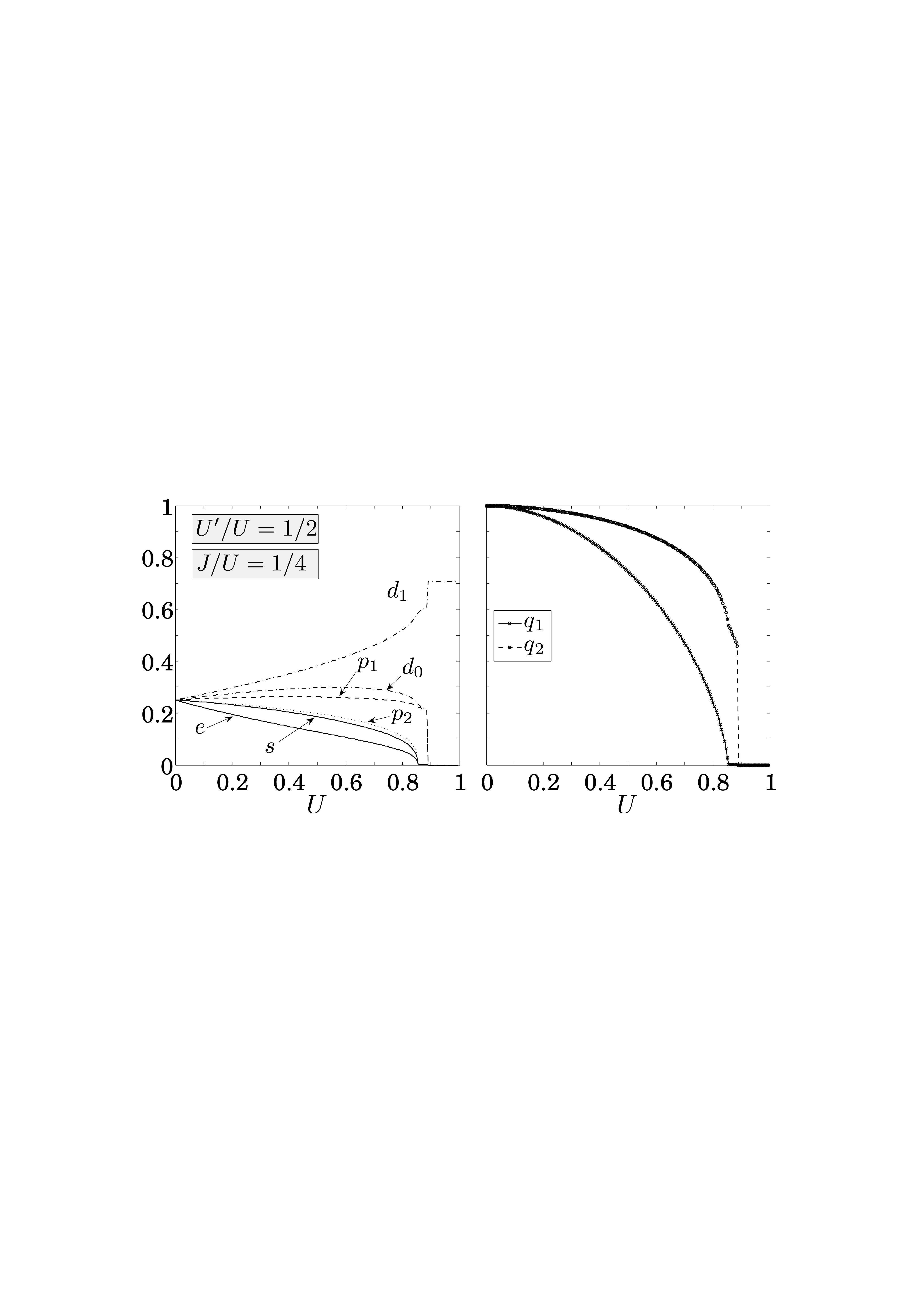}
\caption{Mean fields of the slave bosons and band-renormalization factors $q_{\alpha}$ for $U'/U=1/2$ and $J/U=1/4$.}\label{fig:bosons05}
\end{figure}

\begin{figure} \centering
\includegraphics[width=0.9\linewidth]{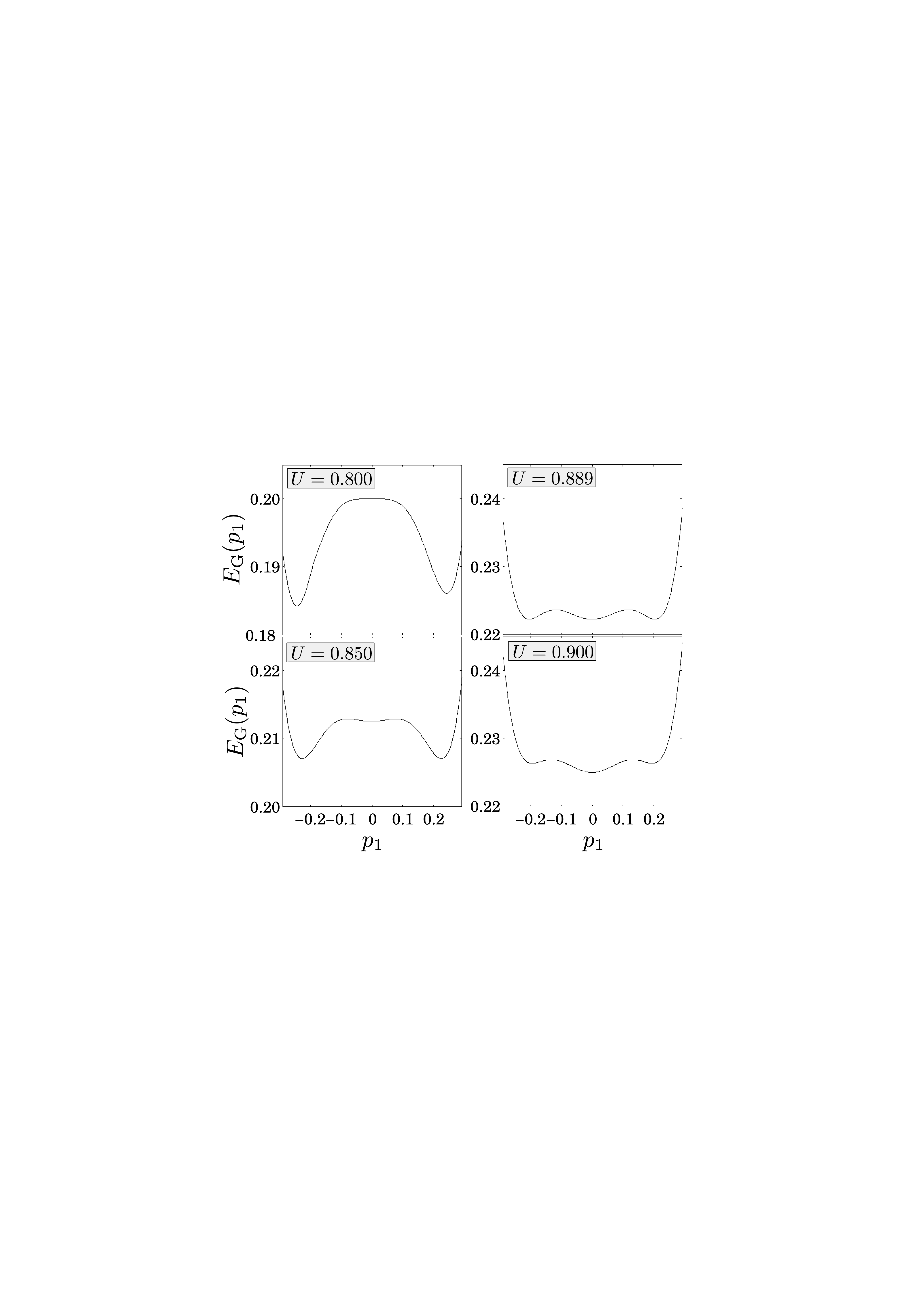}
 \caption{Ground-state energy as a function of
   $p_1$ for different values of $U$, where $U'/U=1/2$ and $J/U=1/4$. $p_1$ is the order parameter for the
   Mott transition in the second band. At $U\approx 0.89$ the metallic
   solution $p_1\approx 0.2$ and the insulating solution $p_1=0$ are
   degenerate. This results in a first order transition.}\label{fig:energy}
\end{figure}

To illustrate the occurrence of the OSMI and the first order transition line
we show in Fig.~\ref{fig:bosons05} the slave-boson mean fields and the
band-renormalization factors for a fixed ratio $U'/U=1/2$ and $J/U=1/4$. We
clearly see that there is a sequence of individual transitions and that $q_2$
jumps at $U\approx 0.89$ from a finite value to zero.  The discontinuity is
also observed in the mean fields $d_0$, $d_1$ and $p_1$. We computed the
ground-state energy as a function of $p_1$ for different values of $U$, where
$U'$ and $J$ have the same ratio as above.  This is shown in
Fig.~\ref{fig:energy}. Note that $p_1^2$ represents the probability to find at
a particular site one electron in orbital 1 and either no electron or two
electrons in orbital 2 and serves therefore as the order parameter for the
Mott transition in the second band.  At $U\approx 0.89$ the metallic solution
$p_1\approx 0.2$ and the Mott-insulator solution $p_1=0$ are degenerate. This
results in a first order transition and a finite jump in $p_1$ and
consequently also in $q_2$ (Fig. \ref{fig:bosons05}).

\subsubsection{Effect of a crystal field}

Until now we have assumed that the two bands are both centered symmetrically
around the Fermi energy. What happens if a crystal field splits the atomic
energy level for the two different orbitals? Let us assume that the overall
system is still half-filled. We introduce an external field
$\eta\sum_i(\hat{n}_{i1}-\hat{n}_{i2})$ in the Hamilton operator
(\ref{eq:tbhm}) which splits the atomic energy levels by $2\eta$.
Particle-hole symmetry allows to concentrate on $\eta\geq 0$.  In the
noninteracting case, this leads to a relative shift of the tight-binding bands
by $2\eta$ and if this shift is bigger than $D_1+D_2$ the lower band is
totally filled whereas the upper band is empty. In this case the system is a
band insulator. How does this band insulator evolve when we turn on the
Coulomb interaction?  Is there a transition from the band insulator to the
Mott insulator, or can we observe a new phase in between?

To answer these questions we investigate the effect of the crystal field on
the mean-field level of the slave-boson approach. In contrast to the case
$\eta=0$ we also keep the spin dependence of the mean fields so as to detect a
possible ferromagnetically ordered state. The external field leads to an
additional term in the variational ground-state energy (\ref{eq:EG}),
$E_{\mathrm{G}}\rightarrow E_{\mathrm{G}}+\eta(n_1- n_2)$.

\begin{figure}
\centering
\includegraphics[width=0.9\linewidth]{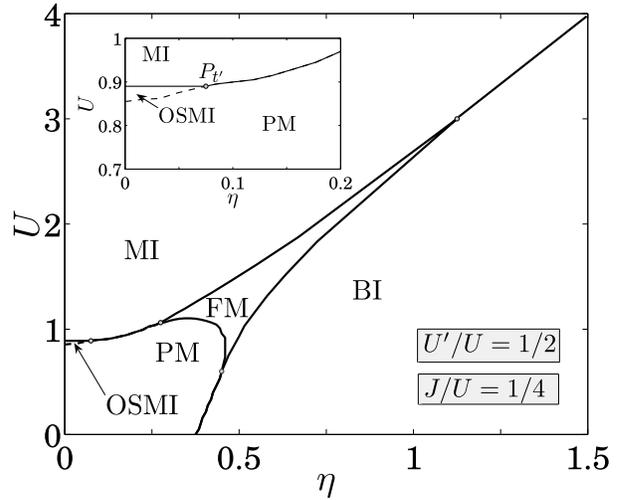}
\caption{Phase diagram in the presence of an external field
  $\eta$ for $U'/U=1/2$ and $J/U=1/4$. The OSMI is limited to a small
  parameter regime as shown in the inset. The crystal field introduces two new
  phases: a band insulator (BI) and a ferromagnetic metal
  (FM).}\label{fig:phasemove}
\end{figure}

\begin{figure}
\centering
\includegraphics[width=0.8\linewidth]{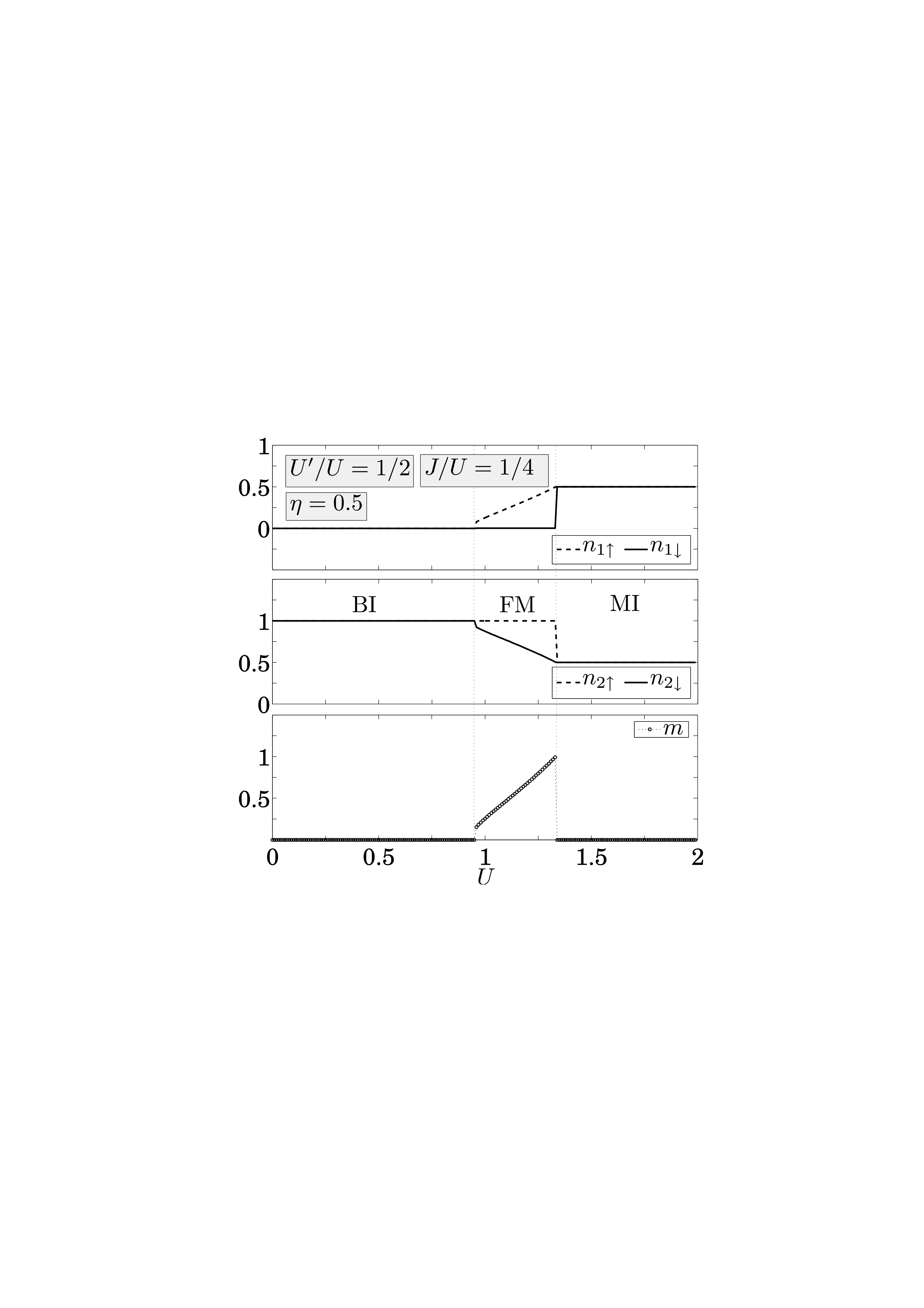}
\caption{The evolution of the charges in the two bands for $U'/U=1/2$ and
  $J/U=1/4$ and the magnetization $m=n_{\uparrow}-n_{\downarrow}$ for fixed
  $\eta=0.5$.}\label{fig:chargesFerro}
\end{figure}

Let us first discuss the phase diagram for a finite Hund's rule coupling.  To
be specific we fix $U'/U=1/2$ and $J/U=1/4$. The result of the minimization of
the ground-state energy for different values of $U$ and $\eta$ is shown in
Fig.\ \ref{fig:phasemove}.
In our slave-boson approach the OSMI phase is restricted to a tiny parameter
regime and only present for small values of $\eta$ as shown in the inset of
Fig.~\ref{fig:phasemove}. The transition in the first band (dashed) is a
second order line which merges the first order transition line (solid) at
$P_{t'}$. If the crystal field is strong enough, the system is in a band
insulating state (BI), i.e.\ one energy band is totally filled whereas the
other is empty. For $\eta>0$ this state is characterized by the mean field
$s_2=1$.  For $U=0$ the transition PM-BI is second-order and happens at
$\eta=(D_1+D_2)/2=0.375$. For a finite $U$ the transition is first-order since
the charge abruptly jumps from $n_{2}<1$ to $n_2=1$. For very strong values of
$U$ and $\eta$ there is a competition between the BI and the MI phase. The
boundary is given by comparing the energy of the BI phase, $U-2\eta$, with the
energy $U'-J$ of the MI phase and yields $U=8\eta /3$ for the above given
ratio of the interaction parameters. The most interesting region of Fig.\ 
\ref{fig:phasemove} lies between these limiting cases where a
ferromagnetically ordered metal (FM) is observed. This state is twofold
degenerated and triggered by the finite Hund's rule coupling. Within our
approximation we always find maximal spin polarization which is characterized
by a finite value of the mean fields $\{s_2,h_{1\sigma}, d_{\sigma\sigma},
p_{2\sigma}\}$. We can get an idea of the physical mechanism by fixing
$\eta=0.5$ and increasing $U$ continuously starting at $U=0$. The evolution of
the charge is shown in Fig.~\ref{fig:chargesFerro}.
At the beginning the Coulomb repulsion is too weak to put electrons in the
upper band and $n_{2\uparrow}=n_{2\downarrow}=1$.  At a critical interaction
strength it is energetically favorable to populate the upper band by a few
electrons of the same spin species and $n_{1\uparrow}$ jumps from $0$ to a
finite value. In the lower band $n_{2\downarrow}$ simultaneously jumps from
$1$ to a value $n_{2\downarrow}<1$. The Pauli principle excludes doubly
occupied orbitals in the upper band which reduces the Coulomb energy and leads
to a ferromagnetic order. In addition, $J$ couples the spin between upper and
lower band and we find the same magnetization in both bands:
$m_1=m_2=m/2$. Note that the critical interaction for the transition to the
FM phase depends on the exact choice of the bare DOS. 
Increasing $U$ further increases $n_{1\uparrow}$ up to 1/2 where we find a
first-order transition from the ferromagnetic metal to the Mott insulator.

\begin{figure}
\centering
\includegraphics[width=0.8\linewidth]{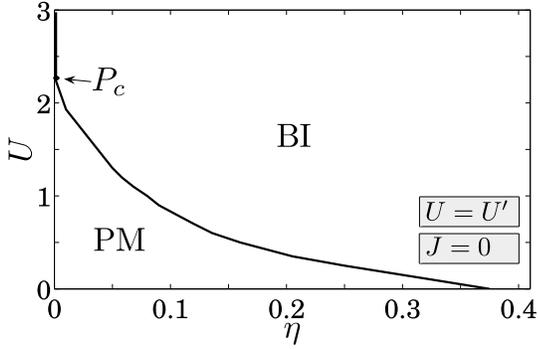}
\caption{Phase diagram in the presence of an external field $\eta$ for $U=U'$
  and $J=0$. At the breakdown of the metallic solution (PM) there is a
  transition to a band insulating phase (BI).}
\label{fig:phasemove1}
\end{figure}
Let us now turn to the case of vanishing Hund's rule coupling $J=0$ and
$U'=U$.  As shown in Fig.~\ref{fig:phasemove1}, a qualitatively different
phase diagram is observed. For $\eta=0$ we saw in Sec.~\ref{subsec:J0} that
the paramagnetic metal is quite stable due to the enhanced degeneracy of the
lowest atomic configurations and that there is a joint transition to the
Mott-insulating phase. This continuous transition is denoted by $P_c$ in
Fig.~\ref{fig:phasemove1}. An arbitrarily small field $\eta$ lowers the energy
of the BI phase compared to the Mott insulator and therefore we find at the
breakdown of the metallic solution for any finite $\eta$ and $U$ a first-order
transition to the BI phase characterized by the mean field $s_2=1$. Similar to
a finite $J$, a finite $\eta$ lifts the degeneracy of the six lowest on-site
configurations, orbital fluctuations are suppressed and therefore the
stability of the metallic phase is reduced with increasing crystal field.

Note that our calculations simplify the true behavior of the system near the
transition lines because the uniform mean-field approximation always
reproduces the results of the atomic limit whenever the kinetic energy
vanishes. Nevertheless they give some insight of the rich behavior of the
system in the present of a crystal field which relatively shifts the atomic
energy levels of the two orbitals.

\subsubsection{Mott transition away from half filling}

\begin{figure}
  \centering \includegraphics[width=0.8\linewidth]{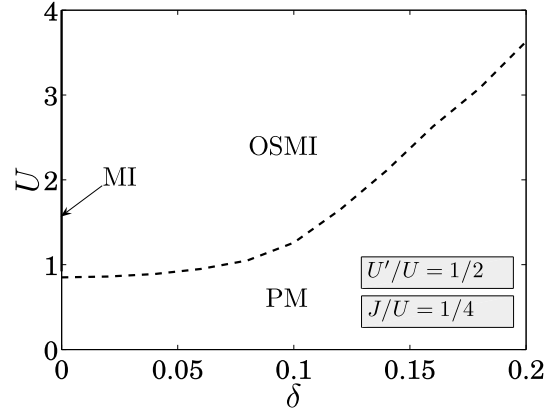}
\caption{Dependence of the Mott transition on the level of doping.}
\label{fig:deltaphase}
\end{figure}

We now address the question of what happens away from half filling,
$n=2-2\delta$, but again with zero crystal-field splitting. Particle-hole
symmetry allows to concentrate on $\delta>0$.  In general we observe a Mott
transition in the narrow band which lies at an increased interaction strength
compared to the case $\delta=0$ (see Fig.~\ref{fig:deltaphase}).
Because the second band is always away from half filling it stays metallic.
\begin{figure}
\centering
\includegraphics[width=1\linewidth]{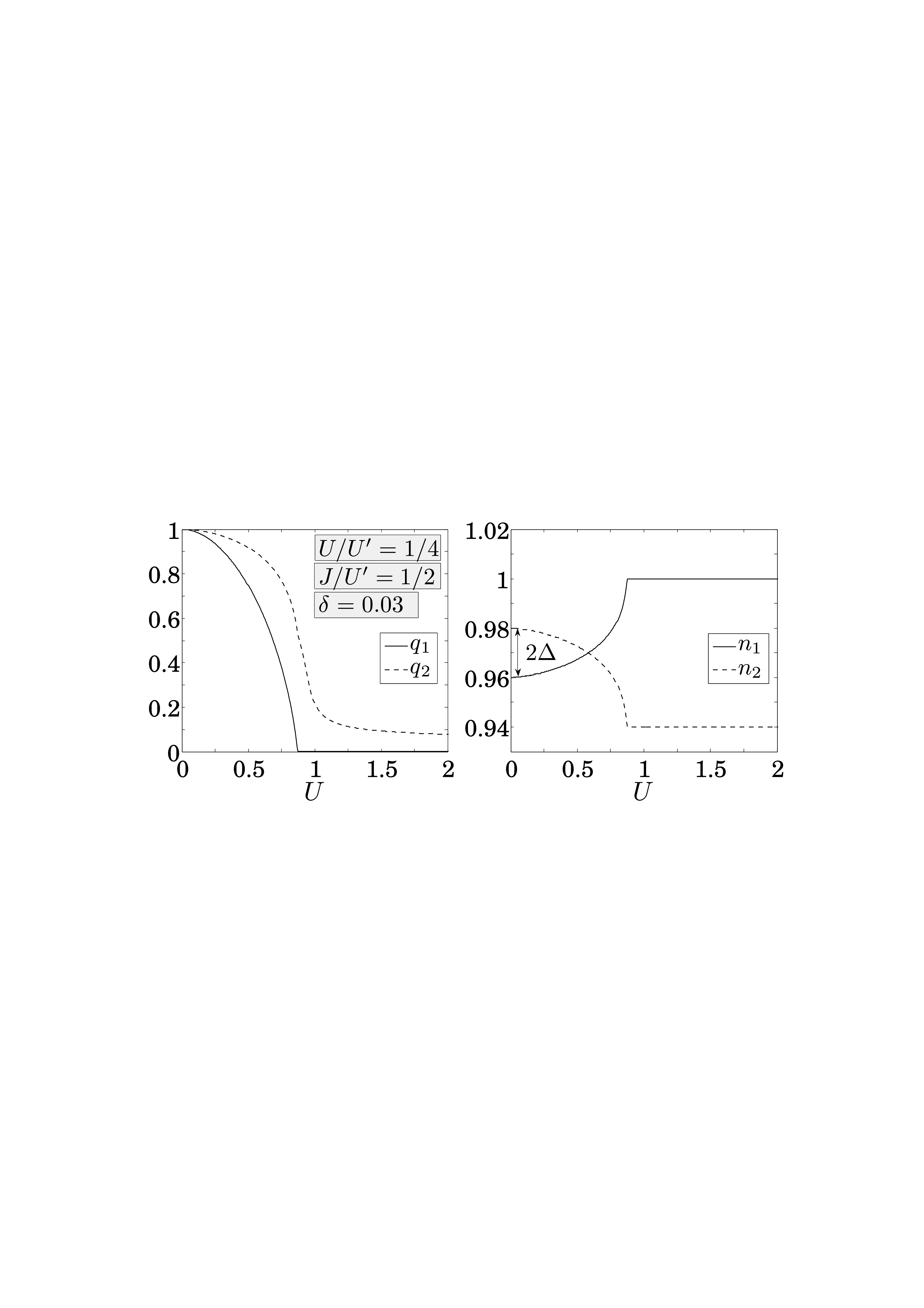}
\caption{Band-renormalization factor $q_{\alpha}$ for
  finite hole doping $\delta=0.03$ and the charge distribution of the two
  bands for a fixed ratio $U'/U=1/2$ and $J/U=1/4$.}\label{fig:delta}
\end{figure}
As a representative example we show in Fig.~\ref{fig:delta} the
band-renormalization factors $q_{\alpha}$ and the charges $n_{\alpha}$ for
fixed ratios $U'/U=1/2$ and $J/U=1/4$ and given doping $\delta=0.03$.  Let us
first look at the noninteracting case, $U=0$. The ground-state energy per site
in this case is
\begin{equation}
  E_{\mathrm{G}}=\bar{\varepsilon}_1^o(1-\delta_1^2)+\bar{\varepsilon}_2^o(1-\delta_2^2)
\end{equation}
where $\delta_{\alpha}$ is the deviation from half filling in band $\alpha$
and $\bar{\varepsilon}_{\alpha}^o$ the average kinetic energy per site in band
$\alpha$ for a half-filled band. Since $\bar{\varepsilon}_{\alpha}^o$ is
proportional to the bandwidth $D_{\alpha}$ we find that the kinetic energy is
optimized by choosing the charge imbalance
$\Delta=(n_1-n_2)/2=(D_1-D_2)/(D_1+D_2)\delta$.  For $\delta=0.03$ and
$D_1/D_2=1/2$ this gives the value $\Delta=-0.01$ as seen in
Fig.~\ref{fig:delta}. Thus, for $U=0$, the narrow band serves as a charge
reservoir that allows to bring the broader band closer to half filling.  With
increasing interaction the Coulomb energy causes a transfer of electrons from
the wide band to the narrow band in order to reduce the intraorbital
repulsion.  Thus, with increasing interaction strength, the broader band
serves as a charge reservoir. This gives rise to a half-filled band
($\Delta=\delta$) at a certain interaction strength and to an OSMT. The
metallic behavior of the second band is due to the finite hole doping and we
find $q_2=2\delta$ up to first order in $\delta$ in the atomic limit
$\bar{\varepsilon}_{\alpha}/U\rightarrow 0$ \cite{Lavagna:90}.

\section{Discussion}
\label{sec:con}
\subsection{Comparison to known results at half filling}
\label{subsec:comp}
On the mean-field level of the slave-boson approach we investigated the Mott
transition in the two-band Hubbard model with different bandwidths and 
confirmed several aspects of previous work.  As reported by several authors
\cite{Liebsch:03, Liebsch:04, Koga:04, Koga:04b, Koga:04c, deMedici:05,
  Ferrero:05, Arita:05, Knecht:05, Liebsch:05, Inaba:05} we observe the
OSMT and consequently an intermediate phase where only the narrow band is
insulating whereas the wide band still has metallic properties.

Our mean-field calculations predict that for $U=U'+2J$ the OSMI phase is
limited to a small parameter regime in the $U$-$U'$ phase diagram which is
characterized by a rather high value of $J$ (Fig.~\ref{fig:phases}).  Compared
to the phase diagram shown in \cite{Koga:04} the strength of the OSMI phase is
strongly reduced. In view of our treatment of the Hund's rule coupling this
can be expected. As pointed out in \cite{Koga:04b, deMedici:05, Liebsch:05,
  Knecht:05} the pair-hopping and the spin-flip term of the full Hund's rule
coupling lead to a stabilization of the OSMI phase. Since these terms are
omitted in our calculations (and also in previous QMC studies
\cite{Liebsch:03, Liebsch:04, Knecht:05}) the OSMI phase is strongly reduced.
Nevertheless, also with an Ising like Hund's rule coupling we can clearly
resolve a sequence of individual Mott transitions in our slave-boson approach.

In addition, we showed that on the mean-field level the PM-OSMI transition is
second-order whereas the OSMI-MI and PM-MI transitions are first-order. Near
these transitions there coexist two different solutions and the energy
crossing results in a first-order transition (Fig.~\ref{fig:energy}).  The
same behavior was reported in \cite{Ferrero:05} in the framework of the
Gutzwiller approximation. Different results were found within other
methods\footnotemark[2]  but non of the used methods is rigorous and the order of
the transitions remains an open problem. Furthermore, temperature as well as
pair-hopping and spin-flip terms might affect the order of the phase
transitions \cite{Liebsch:05}.

For the case $U=U'$ and $J=0$ orbital fluctuations lead to a stabilization of
the metallic phase and for a fixed ratio of the bandwidths $D_1/D_2=1/2$ a
joint second-order transition is observed (Fig.~\ref{fig:qfacsnbosons1}).  For
the case of two bands of much different bandwidths, $D_1\ll D_2$, the
Gutzwiller approximation and related mean-field theories predict the existence
of an OSMI if the ratio $D_1/D_2$ is below a critical value. This was first
reported in \cite{Ferrero:05, deMedici:05}.  DMFT calculations give clear
evidence that the localized band is not a conventional Mott insulator but has
spectral weight down to arbitrarily small energies.  However, an instability
toward an orbitally ordered phase might play an important role for $U=U'$ and
should also be taken into account in future investigations.

\subsection{Results for shifted bands}
\label{subsec:res}
In real materials different atomic orbitals are usually not degenerate so that
each band has non-commensurate filling. This extension of our model has lead
to a considerably richer phase diagram. Such models correspond to the
situation found in $\hbox{Ca}_{2-x}\hbox{Sr}_x\hbox{RuO}_4$ which has three
bands of partial filling. This material has been an initial motivation for the
study of the OSMT \cite{Anisimov:02}.  Anisimov and coworkers proposed that
the Ca-Sr substitution varies band parameters which in the end leads to a Mott
transition in two of the three bands \cite{Anisimov:02}.  A further example
which belongs likely to this class is the compound FeSi. This compound is a
small gap semiconductor \cite{FeSi:02}. On the other hand, replacing Ge for Si
gives rise to a ferromagnetic metal. Alloying FeSi$_{1-x}$Ge$_x$ allows in
principle for a continuous change of the band parameters such that the
transition can be observed. However, the transition is simultaneous
accompanied by an abrupt transition in the crystal lattice \cite{FeSiGe:03}.

Within the mean field approximation we find the following situation. For small
crystal field splitting an OSMT is observed. In general, the Mott transition
is shifted to higher values of the interaction parameters. Due to the crystal
field, a band insulating phase and, in the presence of the Hund's rule
coupling, also a ferromagnetic phase appear in the phase diagram
(Fig.~\ref{fig:phasemove}). In the ferromagnetic phase, a few electrons
populate the upper band with a finite net magnetization.  For the case $U=U'$
and $J=0$ we find a totally different behavior (Fig.~\ref{fig:phasemove1}).
With increasing field, the metallic phase is less stable because the crystal
field suppresses orbital fluctuations, similar to a finite $J$, by breaking
the degeneracy of the local states.

For finite doping our calculations suggest that there is in general a Mott
transition in the narrow band for not too strong doping.  Although strongly
correlated, the second band stays metallic due to the finite doping. This was
also reported in \cite{Koga:04}.

\subsection{Conclusions}
\label{subsec:con}
In summary, the mean-field theory based on the slave-boson approach of Kotliar
and Ruckenstein gives results which are in good qualitative agreement with
DMFT calculations. While we restricted ourselves to density-density
interactions, our discussion provides a simple physical picture in a wide
range of parameters.  The transverse spin coupling and the
on-site interorbital pair hopping had to be dropped for practical reasons.
Nevertheless, we believe that the effects are rather of quantitative than
qualitative nature.

The method used emphasizes the on-site correlation and intersite correlations
remain treated at a minor level only. Thus we have ignored symmetry breaking
instabilities which double the size of the unit cell, such as
antiferromagnetic instabilities or orbital order. These orders depend strongly
on the detailed band structures and coupling topologies. In most of our
discussion, however, we neglected the band structure aspect. Obviously nesting
properties would play a major role in this context. Generic bands without
nesting, however, follow more likely the ''plain'' behavior of the simple flat
density of states models that we discussed here.  It would be interesting in
future studies to extend the scheme by including also band structure effects
and the related ordering phenomena.

\begin{acknowledgement}
  We would like to thank A.\ Koga, T.M. Rice, S. Huber, A. Leuenberger, 
  A.\ Georges and M.\ Ferrero for stimulating discussions. This study has 
  been financially supported by the
  Swiss Nationalfonds and the NCCR MaNEP.
\end{acknowledgement}

\end{document}